\documentclass[]{raa}
\usepackage{deluxetable}
\usepackage{amssymb}            % referee version: for submission
\usepackage{graphicx,times}
\usepackage{natbib}
\usepackage{color}
\usepackage[figuresleft]{rotating}

\providecommand{\bm}[1]{\mbox{\bol
dmath$#1$\unboldmath}}

\begin{document}
%\SetRunningHead{H.-L. Li
%ET AL.}{THE SOLAR-TYPE BINARY OO AQUILAE}
%\Received{~~~~/~~/~~}%{yyyy/mm/dd}
%\Accepted{~~~~/~~/~~}%{yyyy/mm/dd}

\title{Possible Detection of A Flare-associated Coronal Mass Ejection on A M-dwarf}

% \volnopage{ {\bf 2009} Vol.\ {\bf 9} No. {\bf 10}, 1103--1118}
\volnopage{}
	\setcounter{page}{1}

	\author{J. Wang\inst{1,2,*} 
	}
 \institute{Guangxi Key Laboratory for Relativistic Astrophysics, School of Physical Science and Technology, Guangxi University, Nanning 530004, China, {wj@nao.cas.cn}. 
 \and Key Laboratory of Space Astronomy and Technology, National Astronomical Observatories,
Chinese Academy of Sciences, Beijing 100101, China. 
%  \and Miller Institute for Basic Research in Science, University of California, Berkeley, CA 94720, USA\\
%\vs \no
%	{\small Received * ; accepted *}
}

\abstract{
We here report a probable detection of a stellar coronal mass ejection (CME) in active M dwarf
KIC\,8093473 by performing an analysis on its time resolved X-ray spectra observed by XMM-Newton
satellite. Compared to the value at quiescent state and the interstellar one,
our spectral modeling returns a marginal (and probably evolving) excess of hydrogen column density in 
the flare state at a significance level of 1$\sigma$, which can be
understood by an additional absorption due to a flare-associate CME. The CME mass is then estimated 
to be $\sim7\times10^{18}-2\times10^{20}\ \mathrm{g}$ according to the ice cream cone model.
}

 \authorrunning{Wang}            %author_head in even pages
 \titlerunning{Possible CME on A M-dwarf}  % title_head in odd pages
 \maketitle

\section{Introduction}

As an analogy with the Sun, a stellar coronal mass ejections (CME, e.g., Leitzinger \& Odert 2022) being manifested by a
large scale expulsion of the confined and magnetized plasma into the interplanetary space (e.g., Kahler 1992;
Tsuneta 1996; Kliem et al. 2000; Karlicky \& Barta 2007; Li et al. 2016; Jiang et al. 2021) is expected for
solar-like and late-type main-sequence stars.
The study of stellar CMEs is essential for evaluating the habitability of an exoplanet,
especially for nearby M dwarfs (Shields et al. 2016) that are associated with a
habitable zone with a distance less than 0.1AU from the host stars. Theoretically,
the frequent stellar CMEs can tear off most of the
atmosphere of an exoplanet (e.g., Khodachenko et al. 2007; Cerenkov et al. 2017), and compress the magnetosphere, which enables a penetration of energetic particles into
the planetary atmosphere. The subsequent atmospheric chemistry simulations indicate
an enhanced production of greenhouse gas nitrous oxide and HCN in the
atmosphere (e.g., Tian et al. 2011; Airapetian et al. 2016; Barnes et al. 2016).

However, due to a lack of enough spatial resolution, detection of a stellar CME is still a hard task for the current instruments. We refer the readers to Moschou et al. (2019) and
Osten \& Wolk (2016) for reviews on the methods of detection of stellar CMEs. Generally
speaking, a handful of stellar CME candidates have been reported previously according
to either high velocity Balmer line wings or blusehifted emission line in X-ray
(e.g., Houdebine et al. 1990; Argiroffi et al. 2019; Koller et al. 2021; Wang et al. 2021, 2022;
Namekata et al. 2021; Wu et al. 2022; Lu et al. 2022). Other claims of detection of stellar CME include an observed
EUV/X-ray dimming (e.g., Ambruster et al. 1989; Chandra et al. 2016; Veronig et al. 2021) and
a pre-flare dip in an optical light curve (e.g., Giampapa et al. 1982; Leitzinger et al. 2014), although the latter has been argued against since the dip could be also
resulted from an increase in $\mathrm{H^-}$ opacity during a flare.

Additionally, an increased column density revealed in X-ray spectrum, due to an absorption by an
expanding CME, can be used as an indirect CME signature. In fact,
Franciosini et al. (2001) reported that the column density $N_{\mathrm{H}}$ during a flare
is higher than the quiescent value by a factor 5 in a large long-duration X-ray flare
in RS CVn binary system UX Arietis. The temporal decay of the X-ray absorption during a superflare on the eclipsing binary Algol measured by Favata \& Schmitt (1999) was then
explained as a signature of CME by Moschou et al. (2017).

In this paper, we report a probable flare-associated CME from a M dwarf, KIC\,8093473,
according to its increased $N_{\mathrm{H}}$ during a flare revealed by our
time-resolved X-ray spectral analysis, thanks to the
coordinated observations taken by both Kepler and XMM-Newton.
The paper is organized as follows. Section 2 presents a description of the target.
The time-resolved X-ray spectra are analyzed in Section 3.
Section 4 shows the results and discussions.

\section{KIC\,8093473}

Kuznetsov \& Kolotkov (2021) recently matched the \it Kepler \rm (Borucki et al. 2010) and
XMM-Newton (Jansen et al. 2001) simultaneous flare observations, which enables the authors
to successfully identify nine flares in three late type stars, they are KIC\,8093473,
KIC\,8454353 and KIC\,9048551. After examining the raw XMM-Newton data,
we here focus only on KIC\,8093473 simply because one of its X-ray
flare shows enough photon counts for subsequent modeling on time resolved X-ray spectra.

KIC\,8093473 (=2MASS\,J19212847+4355456) is a low-mass ($M_\star=0.274M_\odot$) flaring 
star with an effective temperature of $3357_{-87}^{+171}$ K and a distance of $205.9_{-7.5}^{+8.0}$ pc taken from the Gaia Date Release 2 catalog (Gaia Collaboration et al. 2018). 
A total of 438 flares has been detected in the object from 
the whole long-cadence data (Van Cleve \& Caldwell (2009), Data Release 25, Q1-Q17, 48 months) of the \it Kepler \rm mission by a comprehensive study
carried out by Yang \& Liu (2019). The released energy of the detected flares ranges from $10^{31.7-33.9}\ \mathrm{erg}$.
Since it is above the main sequence (see Figure 1 in Kuznetsov \& Kolotkov 2021),
Kuznetsov \& Kolotkov (2021) argued that KIC\,8093473 with a measured spectral type of M3 (Pecaut \& Mamajek 2013)
is most likely a unresolved binary or a multiple system consisting of M-dwarfs.

\section{Analysis and Results}

We attempt to examine the probable CME feature of the object in this section by
comparing the flaring and quiescent X-ray spectral properties obtained from our 
X-ray spectral modeling. With this motivation, we at first
need to reproduce the X-ray light curve not only in the flaring state, but also in
the quiescent state. In fact, the latter is quite important for extracting a quiescent
spectrum with adequate signal-to-noise ratio.

\subsection{X-ray Light Curves}

The object has been observed in three runs (i.e., Obs.ID=0302150101, 0302150201 and 0600040101)
from 2005 to 2009 by the XMM-Newton onboard EPIC cameras.
However, the observation run of Obs.ID=0302150201 is given up in the subsequent analysis simply
because the source is heavily contaminated by the bad pixel column.

For each of the two remaining observational runs,
the EPIC PN (Struder et al. 2001) data are reduced by the
SAS v19.1 software\footnote{https://www.cosmos.esa.int/web/xmm-newton/} and by the corresponding calibration files. The events
corresponding to patterns 0–4 are selected from the PN data
and the CCD chip gaps are avoided. The bad and hot pixels are
then removed from the original images. The light curve of the source is then
extracted from a circular aperture at the source position with a radius of
15$^{\prime\prime}$. The background light curve is obtained from
a circular source-free region with a radius of 45$^{\prime\prime}$, which is offset
from, but close to, the source.
%The SAS tasks \it epiclccorr \rm is then used to produce background removed light curves.
The SAS tasks \it epatplot \rm and \it epiclccorr \rm
are used for pile-up effect checking and to produce background removed light curves, respectively.

\begin{figure}
   \centering
   \includegraphics[width=10cm]{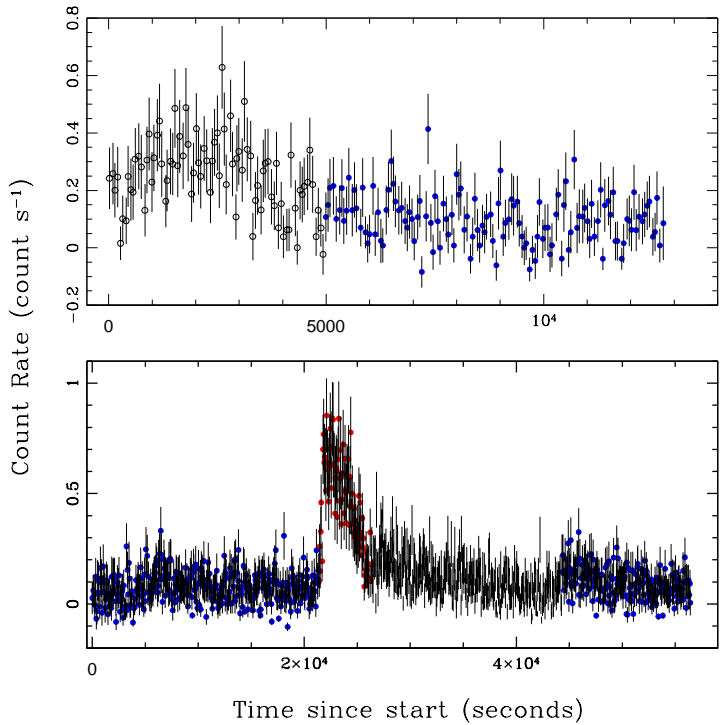}
   \caption{
   XMM-Newton EPIC PN X-ray light curves in 0.2-12 keV energy band. The light curves are
    extracted from two observation runs (\it Upper panel: \rm Obs.ID=0302150101;
    \it Lower panel: \rm Obs.ID=0600040101) with a binning of 60 seconds. The red points mark
    the data used for producing the flaring spectrum, and the blues ones the data for the corresponding
    quiescent spectrum.
}
\label{Fig1}
\end{figure}

\subsection{Flaring and Quiescent X-ray Spectra}

The flaring and quiescent spectra are extracted from the time intervals marked by the solid
red and blue points in Figure 1, respectively. A count rate enhancement at the beginning of the
top light curve, which is probably due to an activity, is excluded from the quiescent spectrum
extraction. Also excluded is the long tail of the flare in the lower light cure.
We extract the source and background spectra by adopting the same regions used in our light curve
extraction. The pile-up effect can be safely excluded after a check by using
the SAS tasks \it epatplot \rm.
The response files needed for subsequent spectral modeling are generated by the task
\it rmfgen \rm and \it arfgen\rm. With the response files, both flaring and quiescent spectra are
grouped by \it grppha \rm in HEASOFT version 6.29. The resulted spectra are shown in the upper
and lower panels in Figure 2 for the flaring and quiescent states, respectively.

\begin{figure}
        \includegraphics[width=\columnwidth]{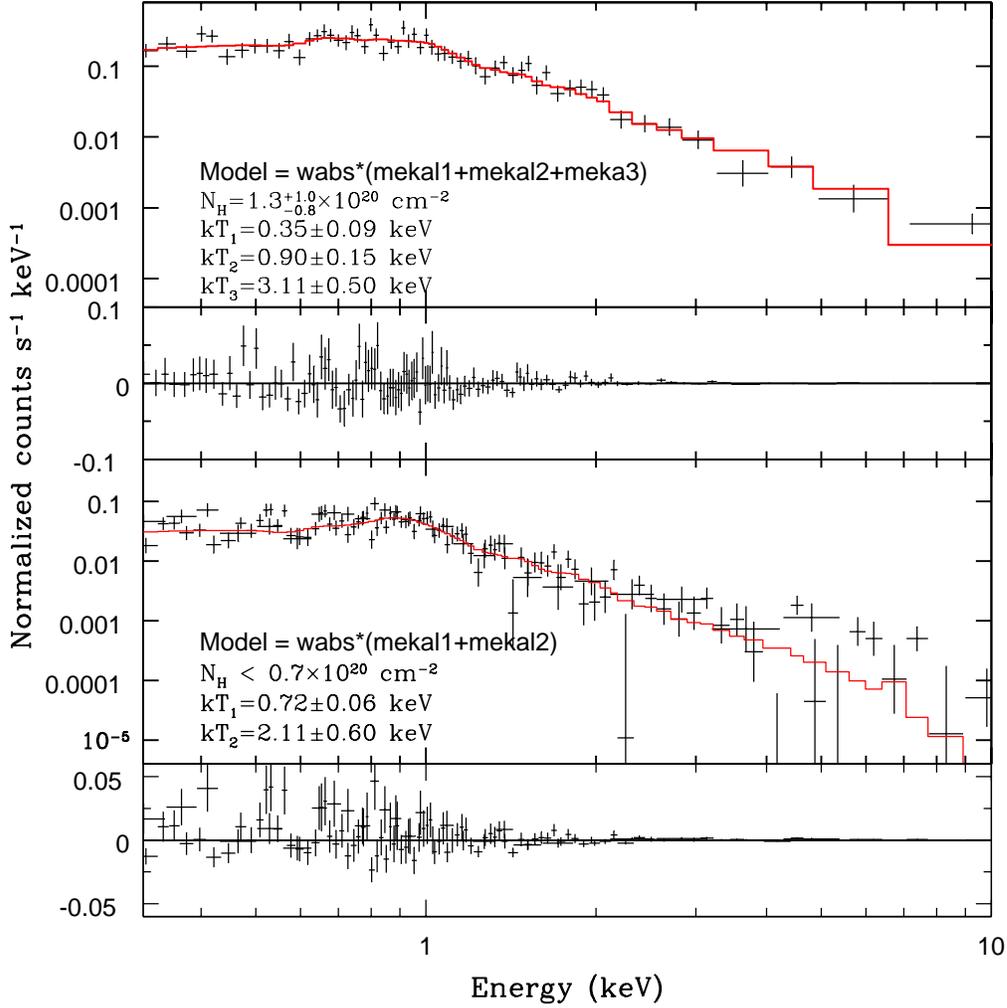}
    \caption{\it upper panel: \rm Flaring X-ray spectra of KIC\,8093473 and the best-fit spectral models expressed as $wabs*(mekal+mekal+mekal)$. The subpanel underneath the spectrum shows the deviations, in units of $\mathrm{counts\ s^{-1}\ keV^{-1}}$, of the observed data from
    the best-fit model. \it Lower panel: \rm the same as the upper one, but for the quiescent state.
    The best-fit model is $wabs*(mekal+mekal)$.
    }
\end{figure}

\subsection{Spectral Modeling}

The aim of this study is to examine the signature of flare-associated stellar CME by comparing the flaring
and quiescent X-ray spectra. The spectra extracted above are then modeled here by XSPEC (v12.11, Arnaud 1996) over the 0.2–10 keV range in terms of the C-statistic (Cash 1979; Humphrey et al. 2009;
Kaastra 2017).

Specifically, we reproduce the flaring spectrum by a linear combination of three optically-thin thermal
plasma with different temperatures, i.e., the MEKAL model emissivities (Mewe et al. 1995).
After taking into account the effect due to an interstellar (or intrinsic) absorption, the model can be
expressed as $wabs*(mekal+mekal+mekal)$.
Although the metal abundance determined from photometry is $\mathrm{[Fe/H]=+0.04 dex}$,
the metal abundance of all the three used MEKAL
components is fixed to be $0.3Z_\odot$ in our spectral fitting since the inadequate count rates.
Most active stars, in fact, are found
to have subsolar coronal metal abundance that is lower than the photospheric value
(e.g., Gudel et al. 1999; 2001; Maggio et al. 2000).
The \rm interstellar column density $N_{\mathrm{H}}$ is set as a free parameter in the fitting.
The quiescent spectrum is fitted by a similar model except for a linear combination of only two
MEKAL components.

The best fits and their parameters are shown in Figure 2 and listed in Table 1. In the table, all of the
quoted uncertainties correspond to a 1$\sigma$ significance level.

\begin{table*}
        \centering
        \caption{XMM-Newton EPIC PN X-ray spectra fit parameters for the flaring and quiescent states of KIC\,8093473.}
        \footnotesize
        \label{tab:example_table}
        \begin{tabular}{lccc} % four columns, alignment for each
                \hline
                  Parameter & Value & Units & Description \\
                  (1) & (2) & (3) & (4) \\
                  \hline
                  \multicolumn{4}{c}{Flare}\\
                  \hline
                  $N_{\mathrm{H}}$ & $1.3_{-0.8}^{+1.0}$ & $10^{20} \mathrm{cm^{-2}}$ & Interstellar column
                  density\\
                  $kT_1$ & $0.35\pm0.09$ & keV & Plasma temperature\\
                  $kT_2$ & $0.90\pm0.15$ & keV & Plasma temperature\\
                  $kT_3$ & $3.11\pm0.50$ & keV & Plasma temperature\\
                  Cash statistics & $64.16/54=1.178$ & \\
          \hline
          \multicolumn{4}{c}{Flare+tail}\\
          \hline
          $N_{\mathrm{H}}$ & $0.9_{-0.7}^{+0.7}$ & $10^{20} \mathrm{cm^{-2}}$ & Interstellar column
                  density\\
                  $kT_1$ & $0.35\pm0.06$ & keV & Plasma temperature\\
                  $kT_2$ & $0.94\pm0.16$ & keV & Plasma temperature\\
                  $kT_3$ & $2.47\pm0.40$ & keV & Plasma temperature\\
                  Cash statistics & $114.16/118=0.967$ & \\
          \hline
          \multicolumn{4}{c}{Quiescent}\\
          \hline
          $N_{\mathrm{H}}$ & $<0.7$ & $10^{20} \mathrm{cm^{-2}}$ & Interstellar column
                  density\\
                  $kT_1$ & $0.72\pm0.06$ & keV & Plasma temperature\\
                  $kT_2$ & $2.11\pm0.60$ & keV & Plasma temperature\\
                  Cash statistics & $82.39/64=1.287$ & \\
                  \hline
        \end{tabular}
\end{table*}

Compared to the quiescent spectrum, the temperature of the hot plasma at the flaring state
increases from $kT\approx2$keV to $\approx3$ keV. Moreover, an additional warm plasma with $kT\approx0.3$ keV
is required to reproduce the observed flaring spectrum. One the one hand, our fitting, in fact, returns
a value of $N_{\mathrm{H}}$ close to zero in the quiescent spectrum. Adopting an interstellar
hydrogen volume density of $0.07\mathrm{cm^{-3}}$ (Paresce 1984) yields a $N_{\mathrm{H}}=0.4\times10^{20} \mathrm{cm^{-2}}$
according to the distance of the star measured by Gaia mission.
This $N_{\mathrm{H}}$ value is consistent with the upper limit given by the modeling,
which suggests an ignorable column density excess at the quiescent state.
One the other hand, the $N_{\mathrm{H}}$ obtained in the flaring state is clearly higher than the
interstellar value estimated above at a 1$\sigma$ significance level, and marginally higher than the modeled
upper limit of $N_{\mathrm{H}}$ at the quiescent state.
A possible explanation of this excess is an additional obscuration due to the material expelled during
the flare, which suggests a flare-associated CME occurred in the star.

\section{Conclusion and Discussion}

We perform a spectral analysis on the time resolved X-ray spectra of active M dwarf KIC\,8093473
observed by XMM-Newton during 2005-2009.
Comparison between the flare and quiescent X-ray spectra enables us to identify a marginal
interstellar column density $N_{\mathrm{H}}$ excess in the flaring state.
A flare-associated CME is a possible explanation of this excess (e.g., Franciosini et al. 2001;
Moschou et al. 2017).

As an additional test, a spectral analysis is performed on the flaring+tail X-ray spectrum that is
extracted from the time interval including both red and black points shown in the lower panel of
Figure 1. The model used in the fitting is the same as that used for the flaring spectrum. The
best fit parameters are listed in Table 1 as well. One can see from the table that the inclusion of
the tail results in a lower average $N_{\mathrm{H}}$, which implies a decrease of $N_{\mathrm{H}}$ with time.
In fact, a decay evolution of
$N_{\mathrm{H}}\propto t^{-2}$ has been revealed in the CME occurred in
the eclipsing binary Algol by Moschou et al. (2017).

We estimate the CME velocity $V_{\mathrm{CME}}$ in terms of the empirical relationship that is established
by Salas-Matamoros \& Klein (2015, see also in Moon et al. 2002; Yashiro \& Gopalswamy 2009)
for solar CMEs:
\begin{equation}
  \log\bigg(\frac{V_{\mathrm{CME}}}{\mathrm{km\ s^{-1}}}\bigg)=(0.20\pm0.08)\log\bigg(\frac{F_{\mathrm{p}}}{\mathrm{W\ m^{-2}}}\bigg)+(3.83\pm0.38)
\end{equation}
where $F_\mathrm{p}$ is the peak soft X-ray flux. By assuming an observation at 1AU from the star,
the measured equivalent peak soft X-ray flux is reported to be $F_{\mathrm{p}}=1.47\ \mathrm{W\ m^{-2}}$ by
Kuznetsov \& Kolotkov (2021). With this value, Eq.(1) yields a
$V_{\mathrm{CME}}=(7.3\pm0.9)\times10^{3}\ \mathrm{km\ s^{-1}}$,
which is much larger than the velocities of $10^{2-3}\ \mathrm{km\ s^{-1}}$ 
observed in solar CMEs (e.g., Yashiro et al. 2004).\rm

We estimate the mass of CME from the measured $N_{\mathrm{H}}$ in the context of the ice cream cone
model (Fisher \& Munro 1984) by following Moschou et al. (2017):
\begin{equation}
  M_{\mathrm{CME}}=\frac{V_{\mathrm{cone}}+V_{\mathrm{hemi}}}{b^{'}-b} N_{\mathrm{H}}\mu
\end{equation}
where $\mu=1.36m_{\mathrm{p}}$ is the mass per proton for gas with solar abundance, $m_{\mathrm{p}}$
is the rest mass of a proton.
$V_{\mathrm{cone}}$ and $V_{\mathrm{hemi}}$ are the volumes of two shells, a hemispherical and
a truncated conical one (see Figure 1 and Equations 1. and 2 in Moschou et al. 2017 for the details).
$b^{'}$ and $b$ are the outer and inner radii of the ice cream part, which can be determined from
the half opening angle $\omega$ and the angular width of the shell $\phi$.

The model depends on the height of the cone from the stellar center, i.e.,
$d=(S+R_\star)/(1+\tan(\omega+\phi))$, where $R_\star$ is the radius of the star and $S$ is
the plasma travel distance from the stellar surface.
The height increases with time as the material is ejected from the stellar surface.
Zic et al. (2015) indicates that in the acceleration phase the dynamical length scale of solar CMEs
have $S\sim15R_\odot$. By taking $S=15R_\star$ and
assuming $\omega=35\degr$ and $\phi=10\degr$, the CME mass could be estimated to be
$(2.1\pm1.9)\times10^{20}\ \mathrm{g}$, after excluding the estimated interstellar hydrogen column density of
$N_{\mathrm{H}}=0.4\times10^{20}\ \mathrm{cm^{-2}}$. The estimated CME mass is close to the high
end of the mass range of a stellar CME compiled by Moschou et al. (2019).
The estimated $M_{\mathrm{CME}}$ is strongly sensitive to the adopted CME length scale.
A much lower mass of $M_{\mathrm{CME}}=(7.3\pm6.5)\times10^{18} \mathrm{g}$ can be obtained if a length scale of $2R_\star$
of a typical magnetic loop is used.

Given the estimated CME mass, Figure 3 shows the location of KIC\,8093473 on the bolometric energy $E_{\mathrm{bol}}$ versus CME mass plot. In addition to quote the other stellar and solar events from
literature directly, we obtain the value of $E_{\mathrm{bol}}=(1.4\pm0.2)\times10^{36}\ 
\mathrm{ergs}$ of KIC\,8093473 from
its X-ray flaring energy ($E_{\mathrm{X}}=1.35^{+0.23}_{-0.20}\times10^{34}\ \mathrm{erg}$)
reported in Kuznetsov \& Kolotkov (2021) by assuming a bolometric correction
of $E_{\mathrm{X}}/E_{\mathrm{bol}}=0.01$ that is typical of the Sun
(e.g., Kretzschmar 2011; Emslie et al. 2012). Similar as the other M-dwarfs reported in Wang et al. (2021)
and Wang et al. (2022),
one can see from the figure that KIC\,8093473 likely deviates from the best fit of the solar data by an
enhanced CME mass.

We argue against a deficient CME kinetic energy in KIC\,8093473 by
naively estimating the kinetic energy of CME along the line-of-sight (LoS) axis
by $E_{\mathrm{k}}
\geq1/2M_{\mathrm{CME}}\overline{\upsilon}^2$, where $\overline{\upsilon}$ is the mean measured LoS velocity.
Combining the CME velocity estimated from Eq. (1) and the CME mass yields
$E_{\mathrm{k}}\approx 2\times10^{36} - 5\times10^{37}\ \mathrm{erg}$. This value is not smaller than
the bolometric energy, and is comparable with the one of $E_{\mathrm{k}}\approx (4.4\pm2.8)\times10^{36}\ \mathrm{erg}$
predicted from the solar relationship $\log E_{\mathrm{k}}=(0.81\pm0.85)+(1.05\pm0.03)
\log E_{\mathrm{X}}$ (Drake et al. 2003). The lack of a deficient CME kinetic energy there
implies a mute of the drag force done by a strong overlying magnetic field
(e.g., Vrsnak et al. 2004; Zic et al. 2015).

\begin{figure}
        \includegraphics[width=\columnwidth]{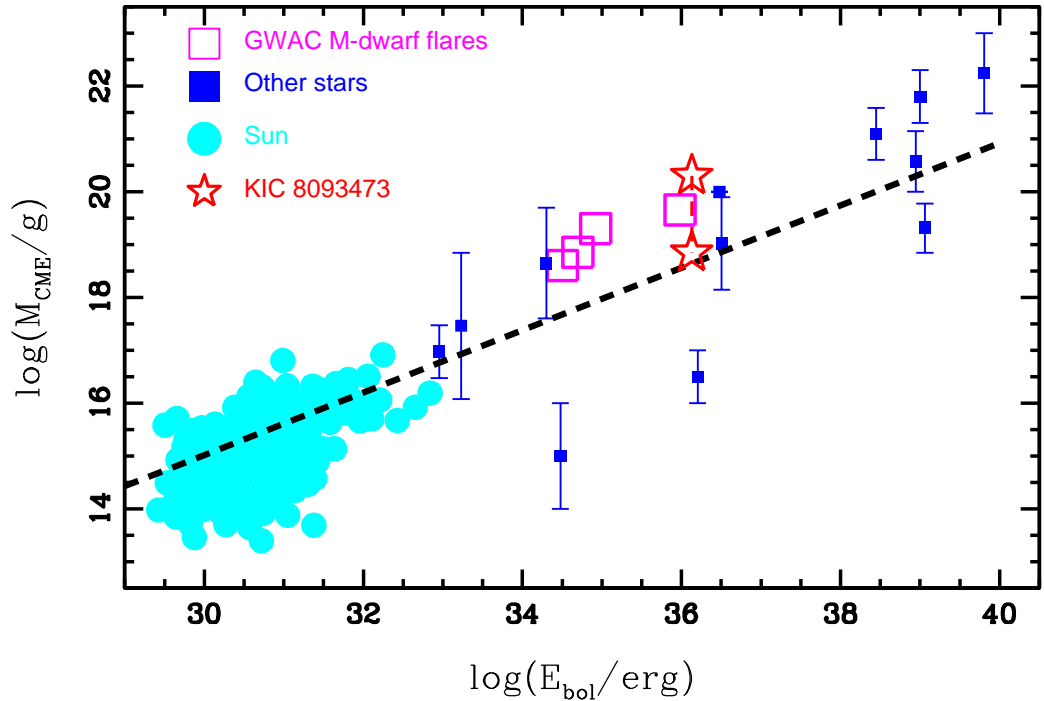}
    \caption{CME mass estimated through different ways plotted against flaring bolometric energy.
    KIC\,8093473 is marked by the two red open stars connected by a dashed line.
    The two points correspond the CME masses estimated by different CME dynamical length scales (see the
    main text for the details).
The four M-dwarf flares studied in Wang et al. (2021) and Wang et al. (2022) are denoted by the magenta open squares. The stellar CME candidates
complied in Moschou et al. (2019) and one studied in Argiroffi et al. (2019) are
shown by the solid blue points. The solar flare-CME events studied in Yashiro \&
Gopalswamy (2009) are denoted by the cyan circles. The best fit to these solar events obtained in
Drake et al. (2013) is presented by the dashed line.
    }
\end{figure}

\begin{acknowledgements}
The authors would like to thank the anonymous referees for his/her careful review and 
helpful comments.
The authors are grateful for support from the National Key Research
and Development Project of China (grant 2020YFE0202100). This
study is supported by the National Natural Science Foundation of
China (Grants No. 12173009), and by the Strategic
Pioneer Program on Space Science, Chinese Academy of Sciences,
grants XDA15052600 and XDA15016500. The author is supported
by the Natural Science Foundation of Guangxi
(2020GXNSFDA238018) and by the Bagui Young Scholars Program.

\end{acknowledgements}

\label{lastpage}

\end{document}